# A Comprehensive Assessment and Benchmark Study of Large Atomistic Foundation Models for Phonons

Md Zaibul Anam,[1] Ogheneyoma Aghoghovbia,[1] Mohammed Al-Fahdi,[1] Lingyu Kong,[2] Victor Fung,[2] and Ming Hu[1,*]

[1]Department of Mechanical Engineering, University of South Carolina, Columbia, South Carolina 29208, USA

[2]School of Computational Science and Engineering, Georgia Institute of Technology, Atlanta, GA 30332, USA

## Abstract

The rapid development of universal machine learning potentials (uMLPs) has enabled efficient, accurate predictions of diverse material properties across broad chemical spaces. While their capability for modeling phonon properties is emerging, systematic benchmarking across chemically diverse systems remains limited. We evaluate six recent uMLPs—EquiformerV2, MatterSim, MACE, and CHGNet—on 2,429 crystalline materials from the Open Quantum Materials Database. Models were used to compute atomic forces in displaced supercells, derive interatomic force constants (IFCs), and predict phonon properties including lattice thermal conductivity (LTC), compared with density functional theory (DFT) and experimental data. The EquiformerV2 pretrained model trained on the OMat24 dataset exhibits strong performance in predicting atomic forces and third-order IFC, while its fine-tuned counterpart consistently outperforms other models in predicting second-order IFC, LTC, and other phonon properties. Although MACE and CHGNet demonstrated comparable force prediction accuracy to EquiformerV2, notable discrepancies in IFC fitting led to poor LTC predictions. Conversely, MatterSim, despite lower force accuracy, achieved intermediate IFC predictions, suggesting error cancellation and complex relationships between force accuracy and phonon predictions. This benchmark guides the evaluation and selection of uMLPs for high-throughput screening of materials with targeted thermal transport properties.

[*]Author to whom all correspondence should be addressed. E-Mail: hu@sc.edu

**Introduction**

The rapid advancement of machine learning potentials (MLPs) have significantly accelerated the speed of atomistic calculations of crystalline materials. These MLPs provide a potentially efficient alternative to the traditional first-principles methods like density functional theory (DFT). Nonetheless, DFT methods are quite accurate compared to experimental results, which is validated by different studies.[1–5] This accuracy of DFT comes at the price of high computational cost and poor scalability with system's size.[6,7] To overcome these limitations, MLP has emerged as a powerful tool, providing high-fidelity modeling of complex and larger systems with comparable accuracy at a fraction of computational cost of *ab initio* methods.[8] Mortazavi *et al.*[9] has shown that, the usage of MLPs in materials science is increasing yearly, and there are several studies which have used MLPs to predict different material properties.[10–16] MLPs utilize selected features which capture the complex atomic configurations to train the model and predict the desired properties. In general, the accuracy of the MLP model's prediction depends on several factors such as the quality of the dataset used for training, choice of descriptors, training algorithms, and others. Schmidt *et al.* showed that the usage of large datasets can improve the accuracy of the MLPs.[17] However, acquiring large datasets through high-throughput DFT calculations for training good MLPs presents a significant computational challenge. Moreover, MLPs are quite proficient in predicting properties for atomic configurations they have been trained on or very similar to what they have seen in training dataset, but the performance degrades when they are used to predict properties for untrained or unseen atomic configurations. Kandy *et al.* has presented a comparison of transferability of MLPs across diverse material configurations, and it verifies the previous statement.[18] Therefore, developing MLPs for a specific crystal structure or a material family to predict relevant material properties is nontrivial. Usually, model input is prepared by representing the local atomic environments in a form suitable for MLPs. Depending on the type of MLP, this may involve either constructing explicit descriptors or allowing the model to learn representations directly from the atomic structure. After a few to several rounds of training/validation/adding new data, the MLP model is finalized with acceptable prediction error and then will be readily deployed to predict properties of new atomic configurations. If some new configurations are not well predicted, the process must be repeated to ensure accuracy. Therefore, developing a MLP for a specific material family with limited number of compositions are usually time, atomic configuration, and property dependent, which can be inconvenient for accelerated high-throughput screening of various materials. A good MLP that performs well on predicting one material property does not simply guarantee it will work very well on another property. Thus, it is not practical to develop many individual MLPs with each just covering a specific material family and predicting a specific material property.

To tackle this issue, a promising avenue involves developing the so-called universal machine learning interatomic potentials (uMLPs) which leverage graph neural networks[19] and equivariant methods[20] to encode atomic interactions more effectively. Several uMLPs have shown very impressive accuracy by Matbench Discovery.[21] In that study, uMLPs such as EquiformerV2[22], SevenNet[23], CHGNet[24], MACE[25], M3GNet[26] showed promising results. Usually, the uMLPs use many datasets from different databases such as Materials Project[27], OMat24[28] and others, and use a significant number of parameters (in the order of sub-million to millions) to train and improve the models. This enables time efficient high-throughput screening and large-scale simulations without requiring retraining for every new material system. Though a lot of work reported high accuracy for uMLPs, there are several drawbacks for these models. Trujillo *et al.* have shown that uMLPs perform poorly in predicting mixing enthalpies for 21 different isostructural binary alloy compared to DFT.[29] Also, Restrepo *et al.* used five different uMLPs to predict several material properties for steels such as vacancy formation energy, surface energies and others, and it is concluded that the pre-trained uMLP models are not yet accurate enough to use in surface or defects related property predictions.[30] This is understandable considering that the majority of the training data used for training those uMLPs are bulk (3D) pristine crystalline systems, i.e., without defects or surface effects. As a result, the trained uMLPs could have some problems in capturing the nature of missing atoms in the neighborhood such as the cases of surface and vacancies.

Thermal transport, governed primarily by phonons (lattice vibration) in crystalline semiconducting and insulating solids, plays a vital role in determining material's thermal conductivity, which is essential for applications ranging from thermoelectric energy conversion to heat dissipation in electronic devices.[31] In recent years, several works have calculated phonons to predict materials for different thermal properties.[32–36] One of the widely used approaches to calculate phonon properties typically involves getting the forces for a displaced atomic configuration in supercells either from DFT or from any MLPs. Then, the atomic forces in the displaced supercells are used to fit the interatomic force constants (IFCs) at different levels of orders, such as harmonic (2$^{nd}$ order) and anharmonic (mainly 3$^{rd}$ order). These IFCs are then used to determine phonon dispersions, group velocities, lifetimes, and lattice thermal conductivity (LTC) with by solving Boltzmann Transport Equation (BTE).[37-42] Another widely used approach to predict phonon transport properties of various materials, including heterogeneous materials and interfaces, is by performing classical molecular dynamics (MD) simulations,[43-46] where the accuracy of force prediction by uMLPs crucially determines the quality of the MD simulations.[47,48] While current uMLPs are effective for energy and force evaluations of small primitive cells (usually less than 10 atoms) across diverse chemical spaces, they often struggle with phonon property predictions due to their inability to inherently capture the interatomic interactions in supercells (containing atoms more than 100 or even 200) needed for accurate IFCs.[49] Moreover, it is imperative for MLPs to achieve a DFT level force accuracy for decent quality phonon property prediction.[50] However, some recent studies that include comparisons of uMLPs have suggested that, only having force accuracy alone is not sufficient, other factors must be considered to achieve reliable phonon properties prediction.[51-54]

Taking all the advantages and disadvantages of the uMLPs into account, in this study, a comparative benchmark study for 1,972 noncubic and 457 cubic structures, which are all acquired from Open Quantum Materials Database (OQMD)[55], is performed using four recently released uMLPs, namely EquifromerV2, MACE, CHGNet, and MatterSim[56]. Regarding the uMLPs to test, we would like to point out that the field of uMLPs, in particular for materials science, has developed very fast in recent years. Lots of new MLPs come out just every few months. At the time this work was initiated, EquifromerV2 models showed promising results in Matbench Discovery and several studies[22,57], and thus in this study, we evaluated the eqV2-L and eqV2-L-DeNS pre-trained models, as well as the eqV2-L OMat MPtrj-sAlex fine-tuned model. Apart from these, only pre-trained models were used for all other uMLPs. Specifically, the MatterSim-V1-5M pre-trained model was employed for MatterSim, CHGNet v0.3.0 was used for CHGNet, and the MACE-MP-0a pre-trained model was used for MACE. The benchmark work for the uMLPs is done in several steps of phonon and related property predictions, and in every step, they are compared with the DFT calculated results. First, displaced supercells for aforementioned structures are used to evaluate the forces using all these uMLPs and comparison is done with the DFT calculated forces. Then, force constant fitting is done up to 3$^{rd}$ order with ordinary least squared (OLS) technique[58] using these evaluated forces and again comparison of both 2$^{nd}$ and 3$^{rd}$ order IFCs is done with those fitted from DFT forces. These force constants are used to calculate LTC utilizing ShengBTE[59] package and again similar analogy is done as before. Finally, for some selected structures, uMLP predicted LTC values are compared with experimental results.

**Results and Discussion**

We first evaluate the uMLPs by feeding the displaced supercells into the selected uMLPs for force prediction. The number of structures used, and models employed in this study are described in the "Computational Details". The force evaluation is divided into two sections based on the type of datasets, namely noncubic and cubic structures. Both types of structures are introduced separately into all six models, and their predicted forces are compared against DFT calculated forces. The performance of the models for atomic forces are evaluated based on their root mean square error (RMSE), as illustrated in

Figure 1. In this study, the eqV2-L model, which trained on OMat24 datasets, is denoted as EquiformerV2(omat), the eqV2-L-DeNS model trained on Material Project (MP) datasets is referred as EquiformerV2(MP_trj), and the eqV2-L OMat MPtrj-sAlex fine-tuned model of EquiformerV2(omat) on MP trajectory and Alexandria[60] datasets is referred as EquiformerV2(FT). From Figure 1, it is evident that EquiformerV2 models are performing better than the other models in terms of RMSE for force prediction, which also aligns with Matbench Discovery ranking. However, the most notable observation for both cubic and noncubic structure is that EquiformerV2(omat), with RMSE of 99.54 meV/Å, outperforms EquiformerV2(MP_trj) and EquiformerV2(FT), with RMSE of 119.66 meV/Å and 122.88 meV/Å, respectively. This shows that EquiformerV2 trained on OMat24 dataset can better represent the OQMD structures used here for the benchmark study as compared to MP trajectory datasets. Also, fine-tuning the EquiformerV2(omat) model on MP trajectory and Alexandira datasets is introducing systematic errors for the selected OQMD structures in this case, resulting in reduced prediction accuracy compared to the original model.

From Figure 1, the CHGNet and MACE pretrained model, with RMSE of 131.76 meV/Å and 130.74 meV/Å, respectively, perform comparably to EquiformerV2(FT) and outperform MatterSim, which has a higher RMSE of 140.93 meV/Å. From Figure 1, it can be observed that the deviation of forces from DFT values is lower for MatterSim and EquiformerV2 models compared to MACE and CHGNet. So, to have a proper comparison of the models, it is essential to consider not only RMSE of the forces but also the overall spread of the force predictions by the models. To tackle this issue, we provide a comparison of predicted forces with DFT for noncubic and cubic structures in Supplementary Information (SI) Figures S2 and S3, and the performance evaluation is done with both RMSE and $R^2$ metrics. The noncubic structures for all models, except EquiformerV2(omat) model, present higher RMSE values in contrast to cubic structures. Also, the $R^2$ values for all models are lower for cubic structures compared to noncubic structures. From Figure S3, we can see that cubic structures have smaller overall force magnitudes (0 – 0.5 eV/Å) clustered around the origin, so even slight deviation in forces from reference values can lower $R^2$ and RMSE. However, noncubic structures exhibit a wider spread of forces, so that the models can capture overall variance more effectively for this study, resulting in higher $R^2$ values. For the noncubic structures, in Figure S2, all EquiformerV2 models show the best results with the lowest RMSE values of 100.73, 113.29, and 120.55 meV/Å, respectively, and highest $R^2$ values of 0.83, 0.86, and 0.82, respectively. Even though the RMSE of MatterSim model is higher than CHGNet and MACE models, it has lower $R^2$ values, meaning it can capture the overall trend in the predicted forces slightly better than the other two models. In contrast to noncubic structures, the performance landscape changes in the cubic dataset. The EquiformerV2 (omat) model continues to outperform all others, maintaining the lowest RMSE (95.09 meV/Å) and the highest $R^2$ (0.78). MatterSim performs slightly better than MACE and CHGNet in terms of $R^2$ (0.51 vs. 0.49 and 0.42), though all three models yield comparable results. These observations suggest that EquiformerV2 models generalize well across both types of symmetry classes, however, other models show reduced performance accuracy in cubic system. Interestingly, there are some structures in both noncubic and cubic systems, for which all six MLP models failed to predict accurately. We have listed the detailed information for all these noncubic and cubic structures in Table S1 in Supplementary Information. It is evident that systems with large unit cell sizes tend to be challenging for all models regardless of their symmetry, which confirms our previous hypothesis. For cubic structures, among the seven cubic structures where all six models underperformed, six are identified as spinel type compounds with general formula of $AB_2C_4$, highlighting the models struggle to learn force relationship with spinel structures.

After evaluating the forces for both cubic and noncubic structures with all six models, now we move forward to see how these forces affect the force constant fitting. Here, we used OLS to do the IFC fitting up to the 3$^{rd}$ order. In OLS technique, the force matrix is constructed by[61]

$$F = Ax$$

where $F$ is the vector of all force components from all displaced configurations, $A$ is the design matrix containing the displacement terms and their symmetry-reduced combinations, and $x$ is the vector of unknown IFC parameters, which can be then solved by minimizing the objective function $\|Ax - F\|_2$. For this reason, we have generated 30 random displaced supercells for each structure and got the corresponding forces from the selected uMLPs. Consequently, with the help of OLS method, we fitted the IFCs up to the third order. We used the Pheasy package which implements the OLS method to do the IFC fitting.[62] For each structure we used predicted forces from 30 displaced supercells to construct the IFCs, where the detailed procedure of the IFC fitting can be found in "Computational Details". It is imperative to say that the accuracy of the IFCs is very important to accurately calculate phonon transport properties such as LTC, which is our final goal in this study.[63] So, after fitting the 2$^{nd}$ and 3$^{rd}$ order IFCs, they are compared with the corresponding IFCs derived from DFT-calculated forces, which are shown in Figure 2 and 3, respectively. For the 2$^{nd}$ order IFCs, we computed the trace of each atom-pair's 3×3 matrix, then averaged the absolute values for each structure. The same procedure was applied to the DFT-fitted 2$^{nd}$ order IFCs, and the results were compared. In the case of 3$^{rd}$ order IFCs, which are represented as 3×3×3 tensors, first the absolute values of the diagonal triplets (such as 1-1-1, 2-2-2, 3-3-3) are taken, then the average of these absolute values are taken for each structure and again compared with the DFT-fitted values processed with the same procedure. Similar to the force evaluation case, the 2$^{nd}$ order IFCs fitted from the EquiformerV2 model predicted forces are performing better than other models, especially EquiformerV2(omat). Although the force prediction performance was poor for the MatterSim and EquiformerV2(FT) models, the quality of their 2$^{nd}$ IFC fitting was surprisingly better than other models in terms of RMSE. A similar trend can also be seen in Figure 3 for 3$^{rd}$ order IFC comparison. However, compared to 2$^{nd}$ order IFCs, the 3$^{rd}$ order IFC fitting accuracy is generally much lower, which is consistent with the findings of Zhou et al.[64] This is because the 3$^{rd}$ order IFCs capture anharmonic interactions, which are more sensitive to small deviations in force predictions. Accurately fitting 3$^{rd}$ order IFCs requires the model to capture subtle variations in forces in different atomic environments and small atomic displacement from equilibrium positions. This also explains the reason for both pre-trained MACE and CHGNet model's accuracy in IFC fitting is the worst for our selected structures in this study. Comparing Figure 2 and 3, we also notice that all models tend to overestimate 3$^{rd}$ order IFCs but this is not the case for 2$^{nd}$ order IFCs. We plausibly explain this as the combined effect of (1) the error in predicted atomic forces; (2) the OLS fitting procedure to get IFCs; and (3) the IFC tensor representation method used here. For example, it is noticed that MatterSim and all three EquiformerV2 models have overestimated atomic forces in the displaced supercells (Figure 1), which could lead to an inflation of the magnitude of 3$^{rd}$ order IFCs while the magnitude of 2$^{nd}$ order IFCs can be more or less preserved when using OLS method to do IFC fitting, and therefore result in large error in 3$^{rd}$ order IFC comparison as shown in Figure 3. The results in Figure 3 point out the grand challenge in accurately predicting phonon anharmonicity of crystalline materials by current MLPs.

An important technical detail regarding the good IFCs performance of the EquiformerV2 model is, we directly used our separate DFT optimized structures to construct displaced supercells and then simply used various uMLPs to evaluate those displaced supercells, i.e., obtaining the atomic forces in the supercells. The atomic forces are then fitted by the OLS method to obtain both 2$^{nd}$ and 3$^{rd}$ order IFCs and subsequently obtain all relevant phonon properties. This process might be different from the finite difference method (FDM) that has been widely used in traditional DFT+BTE approach as well as the recent studies of MLP evaluations.[51-53,65-68] The potential advantage of our fitting method vs. FDM is, the possible cancellation effect for the atomic forces in displaced supercells could lead to higher performance in the quality of final IFCs. This is because the FDM is usually "sensitive" to the accuracy of the force evaluators, i.e., it requires much higher accuracy of the force evaluator, since in FDM approach only one or two atoms are displaced each time in the supercells, and the energy and force difference is very small when performing finite difference calculation, leading to considerable error in IFCs if the force evaluator is not accurate enough. In contrast, all atoms are displaced in our fitting approach and thus the atomic forces differ largely among different configurations. The possible cancellation effect for the atomic forces in displaced supercells could compensate for small errors in predicted forces, resulting in higher performance of fitted IFCs. This could be the reason for different performances of uMLPs evaluated by different groups.

Furthermore, it is worth pointing out that the EquiformerV2 models might have issues with the smoothness of the potential energy surface, which we also noticed very recently during structural optimization. However, we must emphasize that despite such smoothness issue of the potential energy surface, the EquiformerV2 models in particular the EquiformerV2 pretrained model trained on the OMat24 dataset exhibits strong performance in predicting atomic forces and even 3$^{rd}$ order interatomic force constants (IFCs). This implies that the EquiformerV2 model is pretty accurate in predicting the derivatives of potential energy surface with respect to atomic displacements instead of potential energy itself. In this study, we considered direct force prediction method by the uMLPs, and we derived all other parameters like IFCs and all relevant phonon properties from these forces. Therefore, potential energy does not go into our evaluation process at all. We selected several well-known and widely used models, including MACE-MP-0a, CHGNet v0.3.0, as well as models that had received significant attention and demonstrated strong F1 scores on Matbench Discovery at the time this work was initiated, such as MatterSim-V1-5M and several EquiformerV2 models. All models are evaluated with exactly same procedure and benchmarked on the same DFT data, and thus all models received fair comparison.

Dynamic stability is one of the required conditions for high-throughput screening of functional materials. A dynamically stable crystal structure has no imaginary (negative) phonon frequency throughout the Brillouin zone. This condition ensures that the structure resides at at least local minimum (hopefully global minimum) of the potential energy surface and will not spontaneously distort into another configuration.[69] Dynamic stability check is performed after IFCs fitting for all 2,429 structures. It is worth pointing out that, all 2,429 structures are predicted to be dynamically stable when IFCs fitting is done with DFT calculated forces, i.e., all structures are free of negative phonon frequencies in the full Brillouin zone. In this study, we did not perform structure re-optimization by the MLPs before feeding the structures into the models, rather we employed DFT optimized structures. In Figure 4, the number (specifically the percentage) of dynamically stable structures predicted by each model is presented. It is evident from Figure 4 that for all 2,429 structures studied herein the IFC fitting done with the EquiformerV2(FT) model predicted forces is outperforming all other models. Almost 81% of the total structures have been reproduced to be dynamically stable by the EquiformerV2(FT) model. As the accuracy of the IFC fittings is lower for both pre-trained MACE and CHGNet model, it is imperative that a smaller number of dynamically stable materials could be found from these two models. Specifically, only 53.2% of the total structures were predicted to be dynamically stable by the pre-trained MACE model, while the respective percentage is only 48.3% in case of pre-trained CHGNet model. This result is expected since poor performance in predicting 2$^{nd}$ order IFCs directly influences the dynamic matrix, leading to imaginary frequencies, and thus falsely predicts dynamic instability of a structure. As MACE and CHGNet model failed to accurately predict the 2$^{nd}$ order IFCs compared to other models, their capability to predict dynamic stability is significantly compromised, resulting in a lower fraction of structures exhibiting positive phonon dispersion and an increased occurrence of false dynamic instabilities.

As previously discussed, for LTC calculations, we used ShengBTE package for three-phonon BTE calculations. In this study, our goal is to compare total LTC for the selected structures. The ShengBTE package directly provides us with the propagons (denoted as $\kappa_p$) contribution to the total lattice thermal conductivity.[59] To fully capture the total LTC for a structure, we need to consider the off-diagonal contribution of heat-flux operator. Many studies have shown that the off-diagonal term provides a vital contribution to the overall LTC.[70,71] The wave-like off-diagonal or coherence contribution for two-channel thermal transport behavior is less impactful in case of a high phononic thermal conductivity.

Rodriguez et al.[72] has provided a good insight into specific conditions under which this behavior becomes prominent. By extracting the phonon group velocity and lifetime matrices with the heat capacity from ShengBTE output, we then calculate the off-diagonal or coherence (denoted as $\kappa_c$) contribution to the total lattice thermal conductivity, and the required equations are adopted from the previous study of Rodriguez *et al.*[72] In this study, we consider both phononic ($\kappa_p$) and coherence ($\kappa_c$) contribution to the total LTC. Figure 5 shows the total LTC ($\kappa_{Total} = \kappa_p + \kappa_c$) comparison at 300 K for the selected 130 structures between the uMLP models and DFT. From Figure 5, we can see that the prediction accuracy for EquiformerV2 models, particularly EquiformerV2(omat) and EquiformerV2(FT) model, is quite impressive, with MAE of 0.24 log(Wm$^{-1}$K$^{-1}$) and 0.174 log(Wm$^{-1}$K$^{-1}$), respectively. Such low MAE for the EquiformerV2(FT) model is quite competitive as compared to some of the previously reported studies,[73,74] which confirms the superiority of EquiformerV2 models performance for this study. On the other hand, the $\kappa_{Total}$ prediction accuracy for pre-trained CHGNet and MACE model, with MAE of 0.68 log(Wm$^{-1}$K$^{-1}$) and 1.002 log(Wm$^{-1}$K$^{-1}$), respectively, is relatively poor as compared to the EquiformerV2 models, which is evident from Figure 5. MatterSim and Equiformer(MP_trj) exhibited intermediate performance in this study, with MAE of 0.347 log(Wm$^{-1}$K$^{-1}$) and 0.41 log(Wm$^{-1}$K$^{-1}$), respectively, which are quite good compared to pre-trained MACE and CHGNet results. This result is understandable, as LTC values with BTE solution is heavily dependent on the 2$^{nd}$ and 3$^{rd}$ order IFCs,[75] thus the accuracy of the IFCs determines the quality of LTC calculation. As previously discussed, pre-trained CHGNet and MACE model's accuracy in predicting the IFCs is quite low, which subsequently affects the results of $\kappa_{Total}$ prediction. An important point to be noted is that the predicted LTC values are quite accurate even though the predicted accuracy of IFCs are relatively poor. This is primarily because the predictive performance of the IFCs is notably better for selected 50 cubic and 80 noncubic materials. The RMSE of 2$^{nd}$ and 3$^{rd}$ order IFCs are significantly lower for these 130 materials compared to the RMSE computed across the full dataset of 2,429 materials, which is shown in Figures S4 and S5 in SI. Additionally, if we examine the LTC prediction trends in Figure 5 across all models, it is evident that $\kappa_{Total}$ shows larger deviations from DFT results in the low thermal conductivity region compared to the high LTC region. Materials with higher rate of three-phonon interactions show stronger anharmonicity, which increases the phonon scattering, thus lowering the LTC values.[76] These anharmonic interactions are governed predominantly by 3$^{rd}$ order IFCs. Since the accuracy of 3$^{rd}$ order IFCs are lower in this study than 2$^{nd}$ order IFCs, the errors in predicting $\kappa_{Total}$ are significant in the low LTC region.

We now benchmark the uMLPs with more detailed phonon transport properties. Some phonon properties like phonon group velocity are governed by phonon dispersion relations which is influenced by the 2$^{nd}$ order IFC. Phonon group velocity is directly derived from the gradient of phonon dispersion curves, which is obtained from the dynamical matrix constructed from 2$^{nd}$ order IFCs.[77] Speed of sound (V$_s$) is the phonon group velocity near the Γ-point,[77] as at long wavelength limit acoustic phonons exhibit linear dispersion relation which matches the macroscopic speed of sound in the material. So, we can say that V$_s$ is also dependent on the harmonic propagation of the phonons. Similar to this, the mean square displacement (MSD) is the average displacement of atoms due to lattice vibrations (random thermal motion) and is another fundamental descriptor for heat transport in solids. In most of cases especially at not high temperatures, the MSD is dominated by the harmonic properties of the lattice, i.e., it depends on the phonon frequency and eigenvectors derived from the 2$^{nd}$ order IFCs.[72] Thus, a comparison of predicted V$_s$ and MSD with DFT calculated values can give us a good insight into the accuracy of the 2$^{nd}$ order IFCs. We selected 50 cubic and 80 noncubic structures to do this comparison. In Figures 6 and 7, the comparisons for both properties show an impressive accuracy of EquiformerV2(omat) and EquiformerV2(FT) models, which have a similar trend as shown in Figure 2 for the direct comparison of 2$^{nd}$ order IFCs. Out of the three EquiformerV2 models, the prediction performance by the EquiformerV2(MP_trj) model is the worst, similar to the previously discussed results. This model's prediction accuracy is even lower than the MatterSim model as reflected by the RMSE values shown in the figures. We also noticed that the larger MSD values generally have higher prediction errors by uMLPs

Page **7** of **30**

as compared to lower MSD values, in particular for MatterSim, MACE, CHGNet, and EquiformerV2(MP_trj). Higher MSD corresponds to soft phonon modes and soft lattice and is usually induced by complex local atomic environment such as bond hierarchy. The EquiformerV2(omat) and EquiformerV2(FT) models predict MSD very well across the whole range, indicating that these two models are well trained on large numbers of diverse atomic environments.

Phonon lifetime directly impacts the results of the LTC as it is determined by the phonon-phonon scattering process.[75] Phonon mean free path (MFP) is the product of phonon group velocity and lifetime, thus is inherently determined by both $2^{nd}$ and $3^{rd}$ order IFCs. Figure 8 and 9 shows a comparison of phonon lifetime and MFP with DFT calculated results for the same structures selected in LTC comparison. The results for both lifetime and MFP validate the superior performance of EquiformerV2(FT) model's reliable prediction as previously shown in Figure 5 for the LTC results. These results also validate the intermediate performance of MatterSim and EquiformerV2(MP_trj), while also highlighting the worst performance by pre-trained MACE and CHGNet models in predicting phonon lifetimes and MFP. As previously discussed, increased phonon-phonon interaction due to strong anharmonicity increases phonon scattering, and this anharmonicity is strongly related to $3^{rd}$ order IFCs. This high scattering rate causes the phonon to have shorter phonon lifetime, and consequently, reduced MFPs. Figures 8 and 9 clearly show that the predicted values exhibit greater deviations from DFT in the short lifetime and short MFP regions compared to the higher regions, meaning the grand challenge for strong anharmonic systems. This trend is consistent with the above observation that $3^{rd}$ order IFCs are generally predicted with lower accuracy (Figure 3) compared to the $2^{nd}$ order IFCs (Figure 2) for this study, which has directly influenced the phonon properties prediction. This explains the limitations of uMLPs in capturing anharmonic lattice dynamics effectively, in particular for materials with strong intrinsic phonon anharmonicity, which eventually impacts the accuracy of their LTC predictions. The results also show that accurate predictions for phonon properties of strong anharmonic materials would require extremely high accuracy of atomic forces in displaced supercells.

To further present the reliability of the uMLPs for predicting LTCs, we have compared some of the selected material's LTCs with available experimental values. Table 1 shows the LTCs predicted by all uMLPs, DFT, and the experimentally measured values. From Table 1, it is evident that for some materials like BP and KCl, the EquiformerV2 model's predicted LTC values are closer to experimental results, possibly because of the cancellation of errors in MLP predicted $2^{nd}$ and $3^{rd}$ order IFCs. To have a better understanding of the LTC values in Table 1, and to have a better comparison of the LTC values for the predicted uMLPs, we normalized the LTC values for uMLPs and DFT by their respective experimental values and then plotted the comparison in Figure 10. Most of the models' LTC values cluster around the experimental baseline, which is the horizontal $\kappa/\kappa_{exp}=1$ line. From Figure 10, it is evident that the LTC values predicted by the EquiformerV2(omat) and EquiformerV2(FT) models show strong agreement with experimental data, with MAE of 12.923 Wm$^{-1}$K$^{-1}$ and 12.097 Wm$^{-1}$K$^{-1}$, respectively. These values are remarkably close to the DFT calculated LTC results with MAE of 12.024 Wm$^{-1}$K$^{-1}$, highlighting the high fidelity of these models in reproducing experimentally observed LTCs. In some cases, like MgAl$_2$O$_4$, ZrNiSn and VFeSb materials, they are performing better than the DFT calculated results. For Al$_2$ZnO$_4$, the accuracy of prediction is very poor for all models except the pre-trained CHGNet model, where even DFT calculated LTC is deviated further from the perfect agreement line. Apart from this, CHGNet generally underperforms across the whole dataset, often underpredicting the LTCs. To further understand the performance of all the models, performance evaluation is done with MAE values in logarithmic values, so we can have a direct comparison with the results in previous studies. The MAE values for the uMLPs and DFT are done in contrast with the experimental values and are shown as inset in Figure S6 in SI. The MAE values for MatterSim, MACE, CHGNet. EquiformerV2(omat), EquiformerV2(MP_trj), EquiformerV2(FT), and DFT are 0.36, 0.526, 0.905, 0.222, 0.317, 0.213, and 0.211 log(Wm$^{-1}$K$^{-1}$), respectively. From these MAE values, we can reaffirm our earlier conclusion that the EquiformerV2(FT) model's prediction accuracy is on the level of DFT for this study. Similarly, the EquiformerV2(omat) model also demonstrates a comparable level of performance closely matching both EquiformerV2(FT) and DFT results, which aligns with our previously discussed results as presented in Figure 5.

**Conclusion**

This study presents a comprehensive benchmark for the evaluation of the performance of uMLPs for the full phonon properties of selected 1,972 cubic and 457 cubic OQMD structures. For these OQMD structures, EquiformerV2(omat) model trained on OMat24 shows better prediction accuracy of forces than the other two models in the same machine learning framework, namely EquiformerV2(MP_trj) and EquiformerV2(FT), which are trained on or later fine-tuned with different datasets. This shows the importance of the training dataset in terms of prediction accuracy, which agrees with the similar statement made by Loew et al.[51] Though MatterSim model showed impressive performance in the same study by Loew et al.[51] for 10,000 MDR[79], in terms of force prediction, MatterSim performed the worst out of the six models tested in this study. Force prediction accuracy for pre-trained MACE and CHGNet models is on par with EquiformerV2(MP_trj) and EquiformerV2(FT) models. However, the accuracy of the IFCs has decreased significantly for these models. On the other hand, MatterSim model's IFC prediction accuracy is impressive even though it has the worst performance in terms of accuracy of the forces. Performance of the models in predicting the IFCs has an impact on finding dynamically stable materials. As EquiformerV2(FT) model shows superior performance in predicting IFCs, it also reproduces more dynamically stable materials. This performance has also influenced LTC prediction as EquiformerV2(FT) model shows an incredibly high accuracy with MAE only about 0.174 log($Wm^{-1}K^{-1}$), while the pre-trained CHGNet is the worst model in this study in terms of predicting accuracy of the LTC.

In this study, for our OQMD datasets, EquiformerV2(FT) consistently outperforms other models in IFC fitting, dynamic stability identification, LTC accuracy, and other phonon properties. For other models, our benchmark identified clear limitations in generalizing across diverse structural configurations, especially for models that are trained primarily on specific databases like Materials Project trajectories but showed very promising results for OMat24 datasets. Although the EquiformerV2(FT) model is fine-tuned partially using Materials Project trajectories datasets, it showed excellent accuracy in predicting IFCs and LTCs, highlighting the impact of training data selection and model architecture on predictive accuracy, particularly for phonon related properties. Finally, LTC comparison between all uMLP model predictions and experimental results shows that EquiformerV2(FT) model can achieve accuracy to the level of DFT, which is quite impressive as no previous training is done for this study. These results give great insight into the assessed uMLPs in terms of selecting the best model for high-throughput screening for phonon mediated properties such as lattice thermal conductivity, where these models are readily for use and competitive results can be obtained without training the models from scratch. Our results also highlight the importance of dataset selection and their importance in assessing the performance of uMLPs, especially for thermal transport applications.

## Computational Details

*Dataset Preparation*

Four state-of-the-art uMLPs are used to evaluate their predictive accuracy for forces, and indirect prediction of LTCs by IFC fitting and then solving phonon BTE. 1,972 noncubic and 457 cubic structures, in total 2,429 structures, spanning over 63 number of elements in periodic table, are adopted from OQMD database. The occurrence frequency of the constitutive elements in the datasets is shown in Figure 11. We exclude lanthanide and actinide series, and inert gas elements from the dataset. The 1,972 noncubic structures cover a variety of space groups, and the overall statistical distribution of the space group for all 2,429 structures is illustrated in Figure S1 in SI. The cubic bar showing in Figure S1 corresponds to the 457 cubic structures used in this study, and all others space groups correspond to the 1,972 noncubic structures. All structures were re-optimized by DFT with our own computational parameters. Computational details for structure optimization by DFT can be found in our previous studies.[80-85] After structure optimization, supercells were generated by expanding the primitive cell structures with different supercell sizes suitable for the structure. Supercell size is generally determined by fulfilling the following requirements: (1) the lattice parameter of the supercell in different crystallographic directions is more or less the same to a large extent; (2) the total number of atoms in supercells must be at least 80. Most of the finalized supercells contain atoms in the range of 80 to 300. The atoms in the supercells are then displaced randomly in different directions by a constant displacement of 0.03 Å by the PHONOPY[86] package. By doing so, 30 different displaced supercells are generated for each structure, and the atomic forces in the supercells are calculated by the self-consistency DFT. All 2,429 structures have 30 displaced supercells each, making a total of 72,870 datasets, which are used to evaluate the uMLPs in this study.

*Universal Machine Learning Potentials*

After obtaining the dataset, the four uMLPs selected for this study are used for evaluating forces in the aforementioned displaced supercells. Forces are generated by feeding the displaced supercells into the pretrained models of MatterSim, CHGNet, MACE, as well as both pretrained and fine-tuned models for EquiformerV2. In this study, two pretrained models, one fine-tuned model for EquiformerV2 is employed. In this study, two pretrained models, one fine-tuned model for EquiformerV2 is employed. The first pretrained model eqV2-L (large) is trained on almost 100 million OMat24 datasets and and consists of around 153 million trainable parameters.[28] The other pre-trained model eqV2-L-DeNS is trained on Materials Project trajectory dataset (MP_trj), using the same number of parameters.[28] The fine-tuned model eqV2-L OMat MPtrj-sAlex has followed a two-step training process. First, it is pre-trained on the OMat24 or OC20 datasets, and then it is fine-tuned using checkpoint updates from the MP trajectory and Alexandria[60] datasets. Also, we utilize the MatterSim-V1-5M model for MatterSim, which is trained on approximately 17 million structures covering a wide range of atomic configurations, as well as materials sampled across different temperatures and pressures.[56] For CHGNet, we use the most recent pretrained mode, CHGNet v0.3.0, which has been optimized for improved generalization and predictive accuracy. For MACE, we use the MACE-MP-0a (large) pretrained model, which has been trained specifically on the Materials Project dataset.

*IFC Construction and Phonon Property Calculation*

After feeding the datasets to the pretrained and fine-tuned models, the atomic forces are obtained. These displaced supercells with their corresponding atomic forces are then used for IFC fitting up to the third order. As mentioned before, here we used OLS technique to fit the $2^{nd}$ and $3^{rd}$ order IFCs with cutoff distance for $3^{rd}$ order IFC truncated at the third nearest neighbor. A total of 80 dynamically stable structures from the 1,972 noncubic structures and 50 dynamically stable structures from the 457 cubic structures are randomly selected for lattice thermal conductivity (LTC) calculations. LTC is our focus in this study, and ShengBTE package[59] is used for the calculations of three-phonon scattering at 300 K for all selected structures. When solving phonon BTE, the NGRIDS parameter is large enough to ensure the total phonon-phonon scattering channels to be at least $10^8$, such that the 3-phonon scattering process can be well sampled throughout the Brillouin zone.

## Acknowledgements

This work was supported in part by the NSF (award number 2030128, 2110033, 2311202, 2320292) and SC EPSCoR/IDeA Program under NSF OIA-1655740 (23-GC01).

**Table 1.** Comparison of LTCs between MatterSim, MACE, CHGNet, EquiformerV2(omat), EquiformerV2(MP_trj), EquiformerV2(FT), DFT, and experimental values. Some of the experimental LTCs are taken at temperatures near 300 K. The unit of all LTCs in this table is $Wm^{-1}K^{-1}$.

| | MatterSim | Pre-trained MACE | Pre-trained CHGNet | EquiformerV2 (omat) | EquiformerV2 (MP_trj) | EquiformerV2 (FT) | DFT | Experiment |
|---|---|---|---|---|---|---|---|---|
| **NaBr** | 1.52 | 2.45 | 0.85 | 1.90 | 2.44 | 2.07 | 2.51 | ~ 2.9[87] |
| **KCl** | 3.07 | 3.06 | 2.91 | 6.47 | 6.54 | 7.21 | 8.15 | ~ 7.2[88] |
| **RbBr** | 0.23 | 1.13 | 0.72 | 3.36 | 2.12 | 4.36 | 4.44 | ~ 3.8[89] |
| **CsCl** | 1.31 | 0.08 | 0.36 | 0.72 | 0.48 | 0.79 | 0.82 | ~ 1.1[90] |
| **CsBr** | 0.65 | 0.01 | 0.42 | 0.53 | 0.63 | 0.71 | 0.72 | ~ 0.2[91] |
| **MgAl$_2$O$_4$** | 12.30 | 13.10 | 8.15 | 8.64 | 10.09 | 9.45 | 30.17 | ~ 13[92] |
| **SiC** | 427.10 | 505.72 | 26.65 | 374.13 | 171.08 | 355.75 | 486.4 | ~ 490[89] |
| **Al$_2$ZnO$_4$** | 25.03 | 19.72 | 8.22 | 27.49 | 21.98 | 25.07 | 25.46 | ~ 2.4[93] |
| **InAs** | 16.82 | 31.69 | 0.51 | 26.78 | 9.50 | 24.62 | 30.97 | ~ 27[88] |
| **BP** | 440.39 | 409.21 | 2.64 | 397.65 | 305.64 | 401.70 | 532.8 | ~ 400[94] |
| **NaF** | 22.35 | 24.18 | 14.83 | 21.83 | 21.05 | 22.67 | 24.49 | ~ 16.5[95] |
| **CdS** | 17.33 | 8.36 | 0.55 | 15.23 | 13.87 | 15.68 | 24.20 | ~ 18[96] |
| **GaAs** | 35.86 | 15.15 | 0.18 | 20.09 | 13.87 | 17.84 | 40.07 | ~ 55[89] |
| **NaI** | 0.49 | 2.66 | 0.31 | 1.14 | 1.01 | 1.12 | 1.36 | ~ 1.33[97] |
| **NaH** | 17.48 | 2.83 | 0.59 | 12.79 | 8.80 | 14.30 | 14.45 | ~ 5[98] |
| **VFeSb** | 11.17 | 1.91 | 0.65 | 12.64 | 3.59 | 9.80 | 21.44 | ~ 8[99] |
| **ZrNiSn** | 16.16 | 2.37 | 2.88 | 10.47 | 2.91 | 7.91 | 15.13 | ~ 7.4[100] |
| **AlAs** | 123.50 | 53.23 | 9.68 | 114.41 | 33.06 | 92.64 | 95.36 | ~ 80[101] |
| **CaF$_2$** | 9.78 | 15.04 | 2.99 | 9.20 | 9.68 | 8.65 | 9.22 | ~ 9.7[102] |
| **KBr** | 0.24 | 0.27 | 0.19 | 1.38 | 0.82 | 1.45 | 1.42 | ~ 3[103] |
| **NaCoO$_2$** | 4.43 | 0.04 | 2.05 | 10.87 | 6.24 | 5.87 | 29.80 | ~ 19[104] |

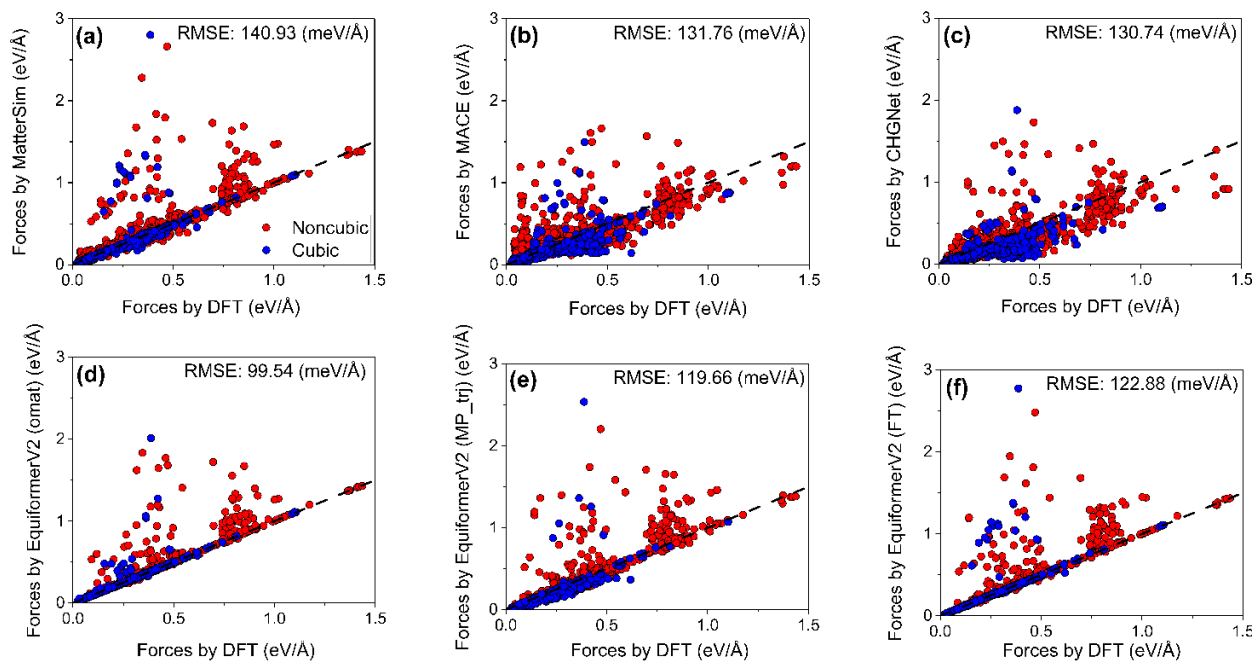

**Figure 1.** Comparison of predicted atomic forces with DFT calculated forces for 2,429 structures using six different uMLPs. Each subplot (a)–(f) corresponds to a different model, with RMSE values reported in meV/Å. The black diagonal line represents the perfect correlation between MLP prediction and DFT calculation. Red and blue data points indicate noncubic and cubic force predictions, respectively. Models evaluated include (a) MatterSim, (b) MACE, (c) CHGNet, (d) EquiformerV2(omat), (e) EquiformerV2(MP_trj), and (f) EquiformerV2(FT). Lower RMSE values indicate better agreement with DFT calculated forces.

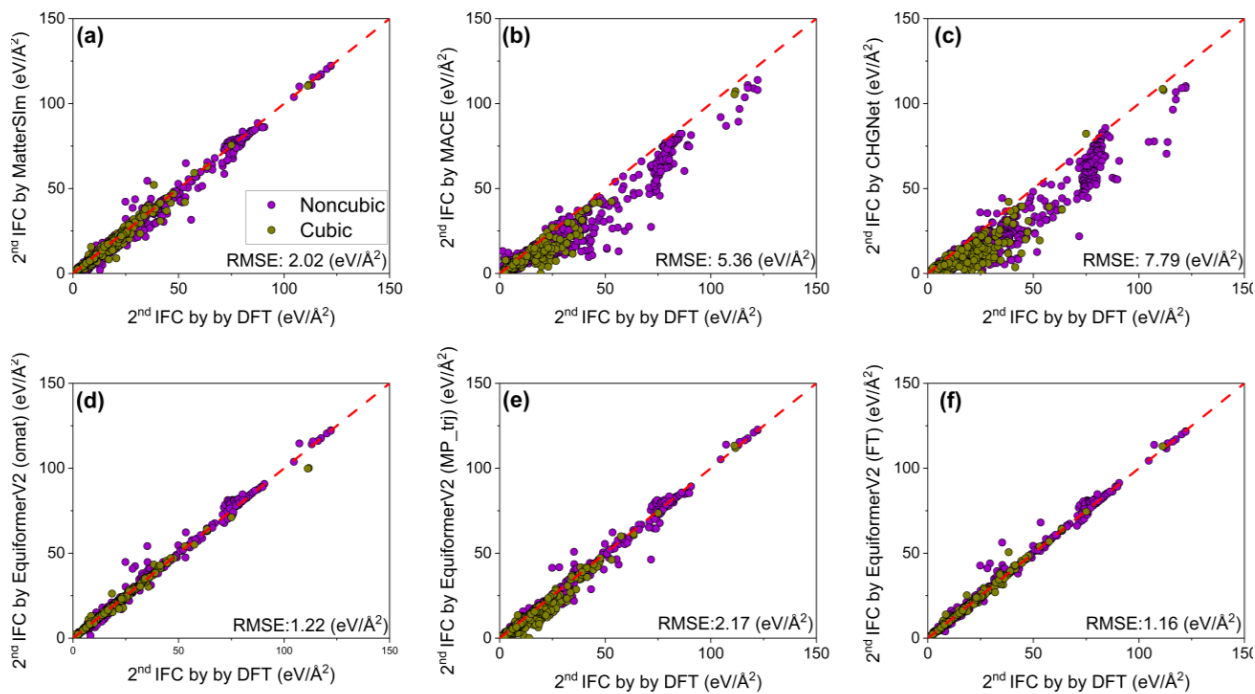

**Figure 2.** Comparison of 2$^{nd}$ order IFCs fitted using forces predicted by different uMLP models with 2$^{nd}$ order IFC constants obtained from DFT calculated forces. Each subplot (a)–(f) represents a different uMLP model: (a) MatterSim, (b) MACE, (c) CHGNet, (d) EquiformerV2(omat), (e) EquiformerV2(MP_trj), and (f) EquiformerV2(FT). The red dashed line represents perfect correlation between uMLPs and DFT-derived force constants. RMSE values indicate the deviation of the uMLP-fitted force constants from the DFT reference. Data points are color-coded based on structure type, with cubic (olive) and non-cubic (purple) structures shown separately.

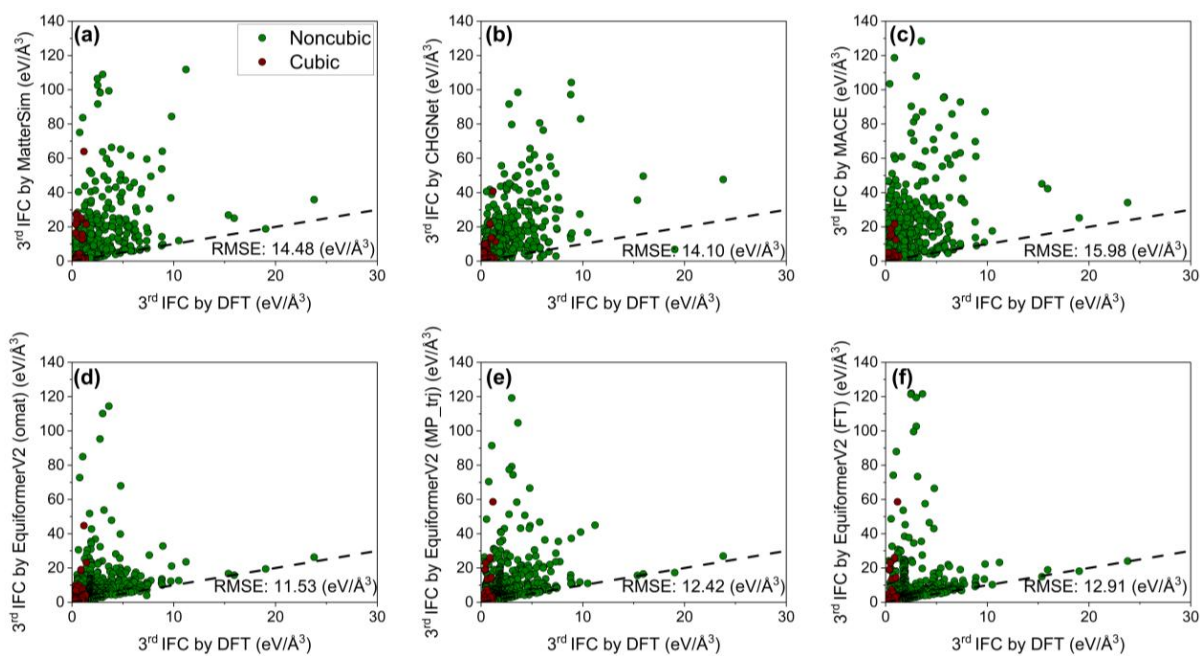

**Figure 3.** Comparison of 3rd order IFC fitting using forces predicted by different uMLP models with 3rd order force constants obtained from DFT-calculated forces for all 2,429 structures. Each subplot (a)–(f) represents a different uMLP model: (a) MatterSim, (b) MACE, (c) CHGNet, (d) EquiformerV2(omat), (e) EquiformerV2(MP_trj), and (f) EquiformerV2(FT). The red dashed line represents perfect correlation between uMLPs and DFT-derived force constants. RMSE values indicate the deviation of the uMLP-fitted force constants from the DFT reference. Data points are color-coded based on structure type, with cubic (wine) and non-cubic (dark green) structures shown separately.

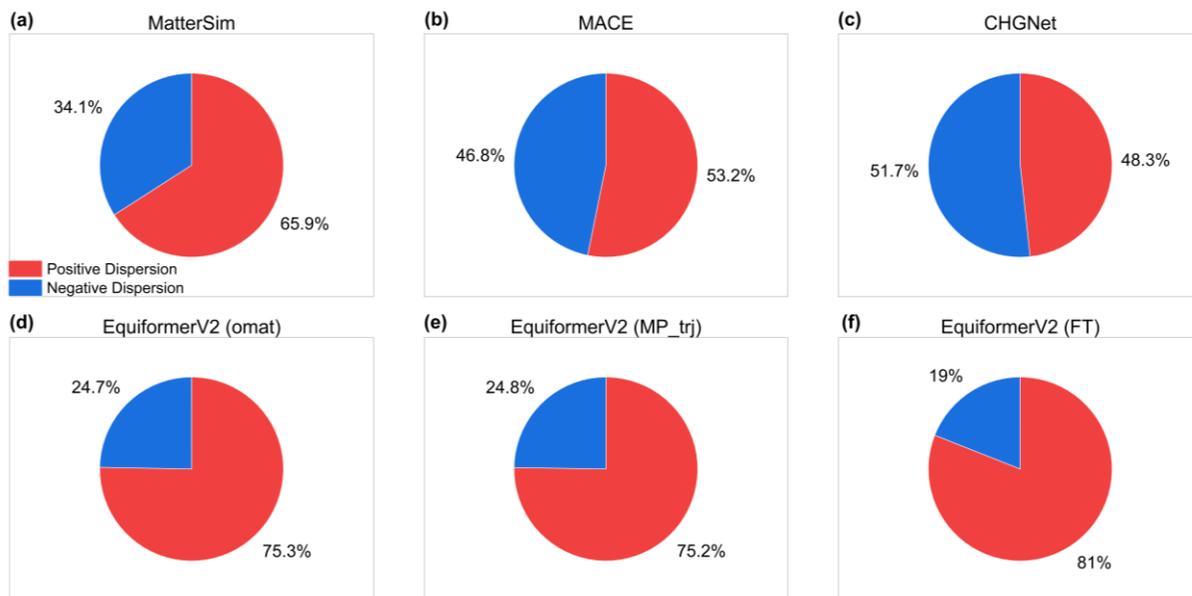

**Figure 4.** Positive and negative dispersion in cubic and non-cubic structures for different uMLP models. Each pie chart represents the fraction (percentage) of structures exhibiting positive (red) and negative (blue) dispersion, which serves as an indicator of dynamic stability. Subplots (a)–(f) correspond to different uMLP models: MatterSim, MACE, CHGNet, EquiformerV2 (omat), EquiformerV2(MP_trj), and EquiformerV2(FT). A higher fraction of positive dispersion structures suggests a higher number of dynamically stable materials are reproduced by a given uMLP model.

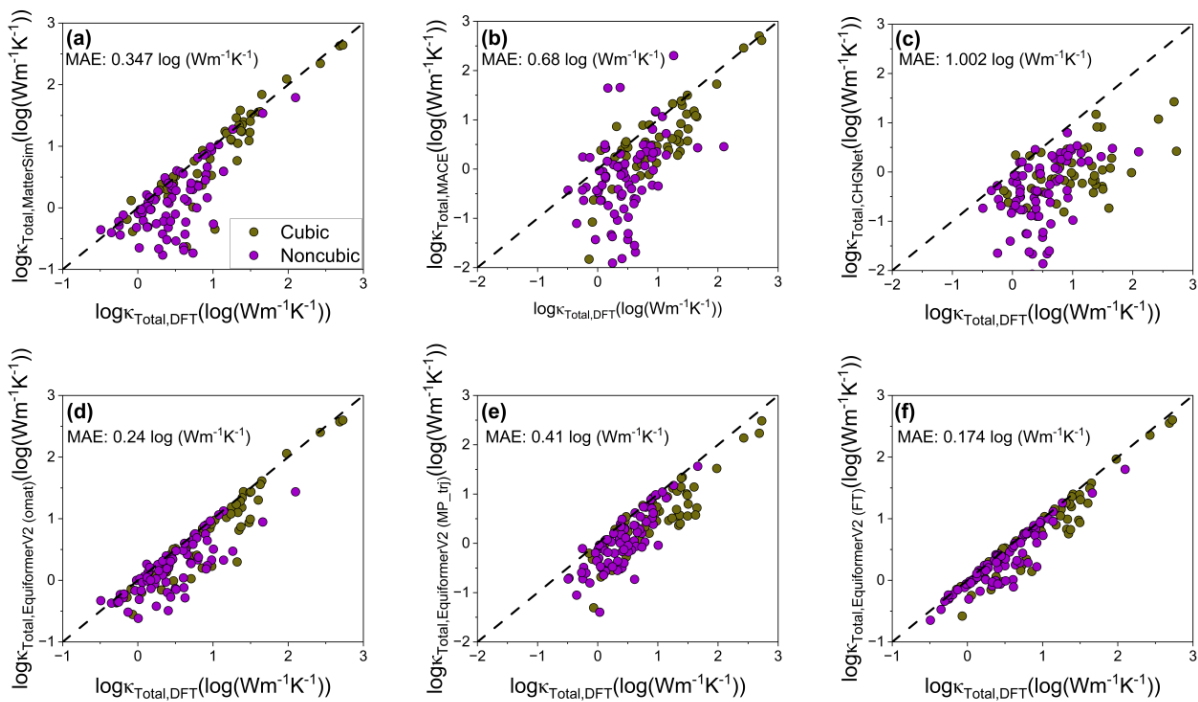

**Figure 5.** Comparison of predicted total lattice thermal conductivities ($\kappa_{total}$), including both phononic and off-diagonal contribution, obtained from uMLPs against DFT-calculated reference values in logarithmic scale for selected 130 structures. Panels represent predictions from (a) MatterSim, (b) MACE, (c) CHGNet, (d) EquiformerV2 (omat), (e) EquiformerV2(MP_trj), and (f) EquiformerV2(FT). Cubic and noncubic materials are presented by olive and purple points, respectively. The dashed diagonal line indicates perfect correlation. Mean Absolute Error (MAE) values, calculated on a logarithmic scale, are shown in each panel to show the MLP prediction accuracy.

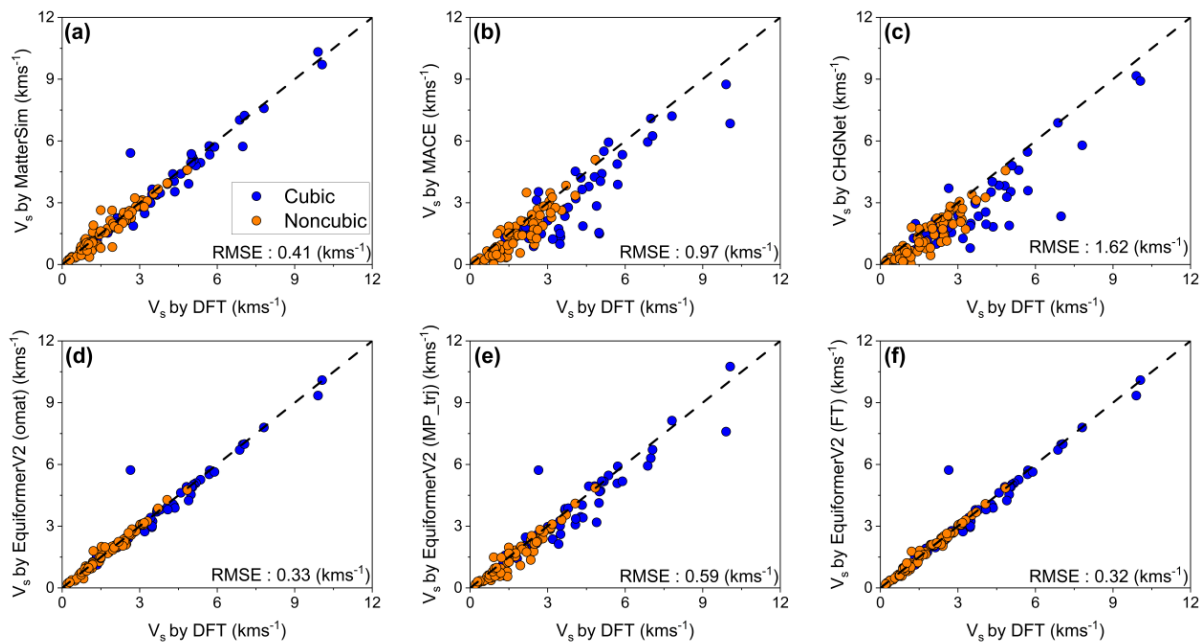

**Figure 6.** Speed of sound ($V_s$) comparison for six selected models with DFT calculations: (a) MatterSim, (b) MACE, (c) CHGNet, (d) EquiformerV2(omat), (e) EquiformerV2(MP_trj), and (f) EquiformerV2(FT). The blue points represent cubic structures, and the orange points represent the noncubic structures. The black dashed line shows the perfect correlation. Performance evaluation is done based on RMSE value pointed in the corner of each plot.

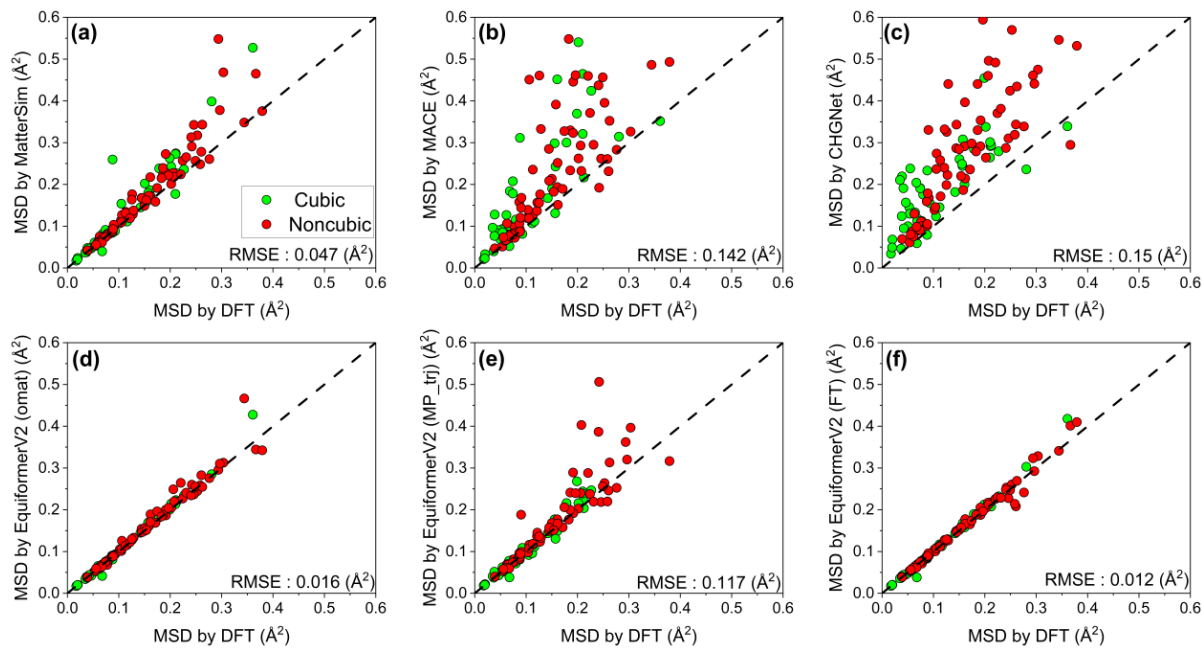

**Figure 7.** Comparison of mean squared displacement (MSD) for six selected models with DFT calculations: (a) MatterSim, (b) MACE, (c) CHGNet, (d) EquiformerV2(omat), (e) EquiformerV2(MP_trj), and (f) EquiformerV2(FT). The green and red points represent cubic and noncubic structures, respectively. The black dashed line shows the perfect correlation. Performance evaluation is done based on RMSE value pointed in the corner of each plot.

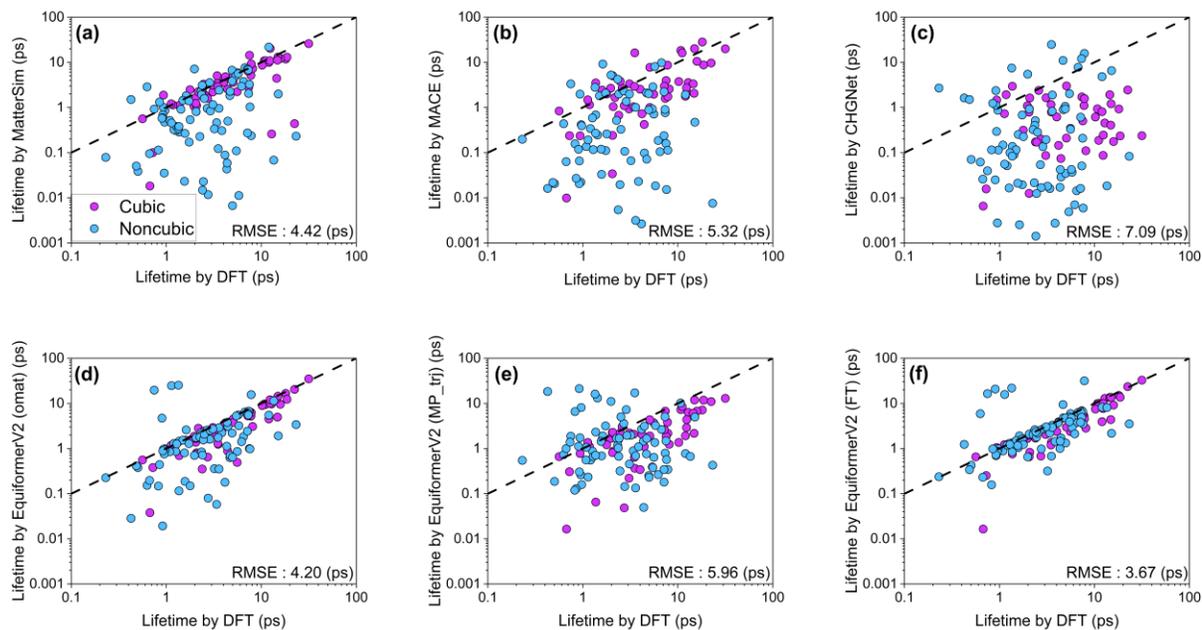

**Figure 8.** Comparison of predicted phonon lifetimes versus DFT calculations for various uMLP models. Each subplot represents a different model: (a) MatterSim, (b) MACE, (c) CHGNet, (d) EquiformerV2(omat), (e) EquiformerV2(MP-trj), and (f) EquiformerV2(FT). The dashed black line represents perfect correlation between predicted and DFT calculated values. Cubic and noncubic structures are shown in magenta and blue colors, respectively. The performance evaluation is done in RMSE value shown in the corner of each subplot. Logarithmic scale is used for better visualization of the data.

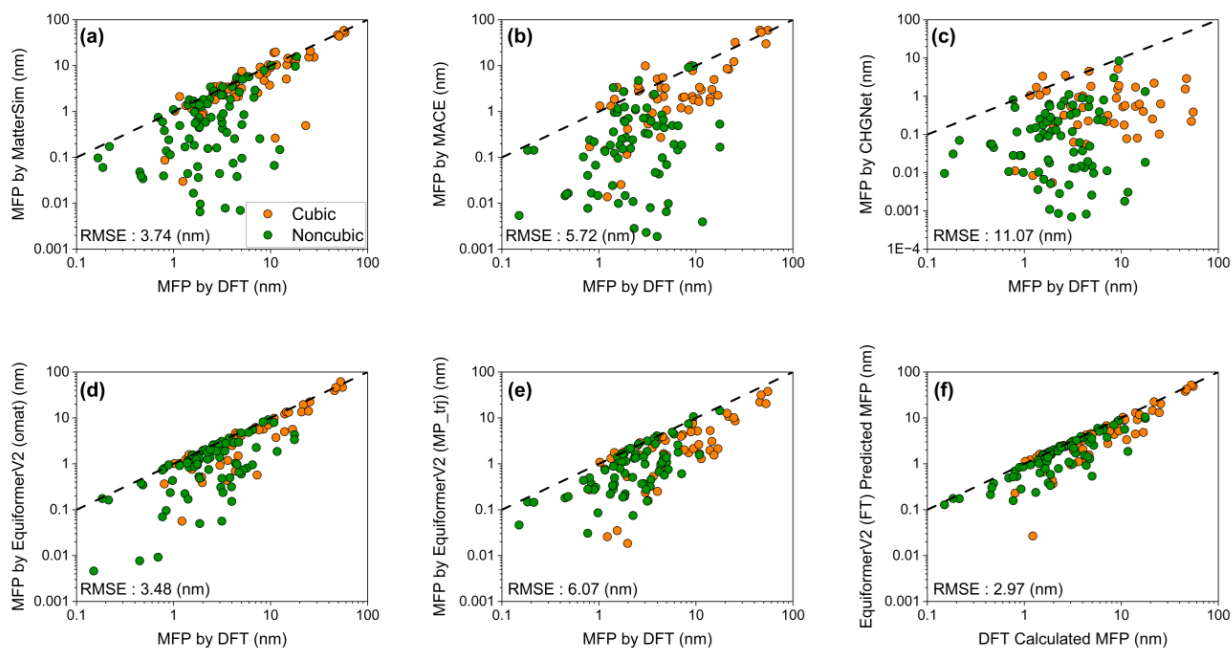

**Figure 9.** Comparison of the average phonon mean free path (MFP) at 300 K predicted by six models with DFT calculations: (a) MatterSim, (b) MACE, (c) CHGNet, (d) EquiformerV2(omat), (e) EquiformerV2(MP_trj), and (f) EquiformerV2(FT). Orange and green points represent cubic and noncubic structures, respectively. The dashed line represents the perfect correlation for MLP prediction and DFT results. Logarithmic scale is used for better visualization of the data.

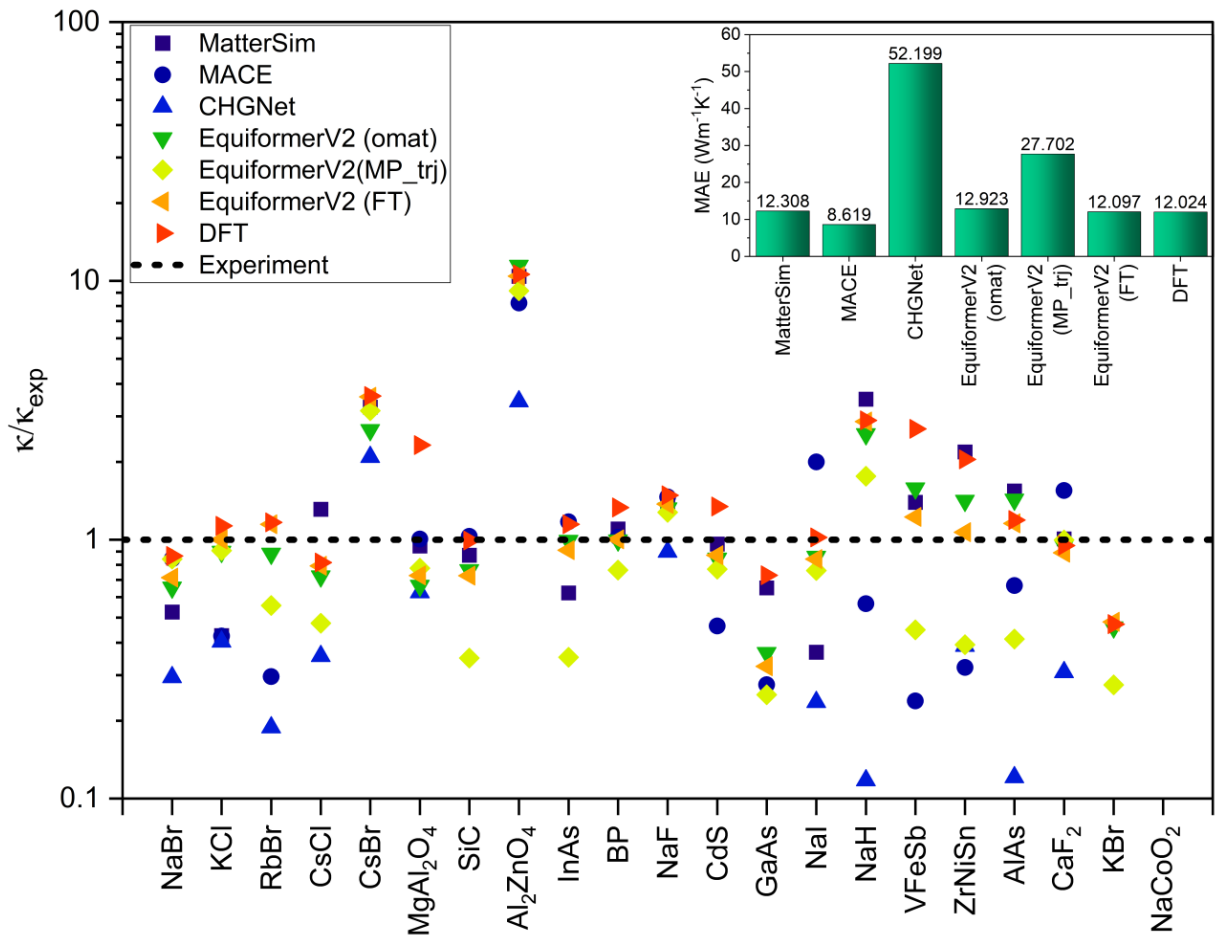

**Figure 10.** Normalized total lattice thermal conductivity (LTC) by experimental values for 22 selected materials. $\kappa$ represents LTCs predicted by MatterSim, MACE, CHGNet, EquiformerV2(omat), EquiformerV2(MP_trj), and EquiformerV2(FT) model and DFT calculated results. $\kappa_{exp}$ represents experimentally measured data. $\kappa/\kappa_{exp}$ is the ratio of LTCs for all selected models with experimental values. The horizontal dashed line at $\kappa/\kappa_{exp}=1$ represents perfect agreement with experimental measurements. The y-axis is shown on a logarithmic scale to better visualize deviations for different models. The inset displays the MAE for each MLP model presenting the prediction accuracy.

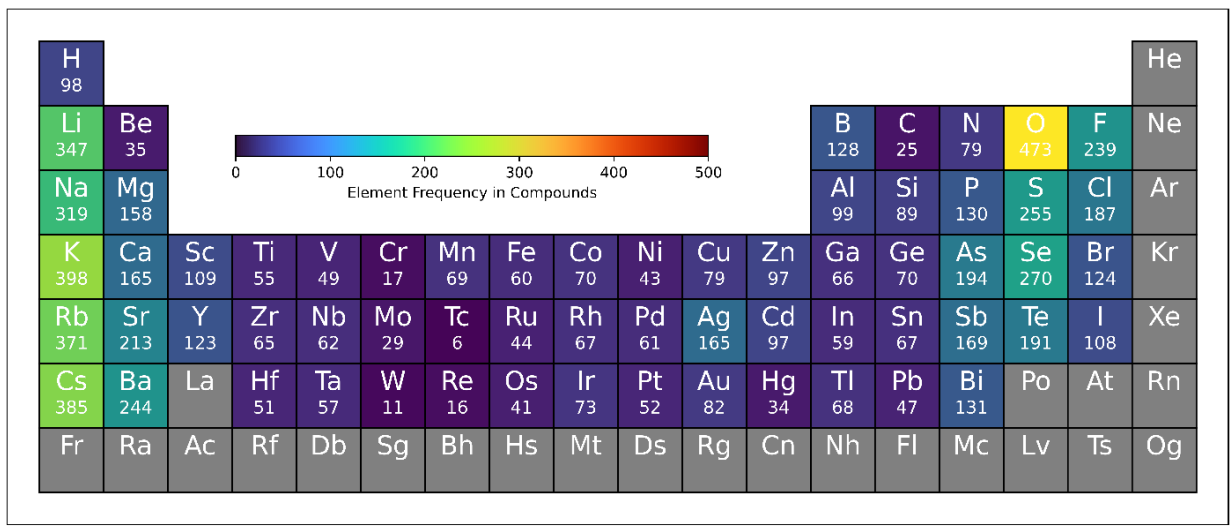

**Figure 11.** Element occurrence frequency for all 2,429 structures benchmarked in this study across the periodic table. Numbers inside the element blocks represent frequency of elements shown in the 2,429 compounds, which is also represented by color coding the whole periodic table. Gray color coded elements in the periodic table are not considered in this study.